# K-space and Image Domain Collaborative Energy-based Model for Parallel MRI Reconstruction

Zongjiang Tu[1], Chen Jiang[1], Yu Guan[1], Shanshan Wang[2], Jijun Liu[3], Qiegen Liu[1]

[1]Department of Electronic Information Engineering, Nanchang University, Nanchang 330031, China.
[2]Shenzhen Institutes of Advanced Technology, Chinese Academy of Sciences, 518055, China.
[3]School of Mathematics, Southeast University, Nanjing 210096, China.

*Abstract*—**Decreasing magnetic resonance (MR) image acquisition times can potentially make MR examinations more accessible. Prior arts including the deep learning models have been devoted to solving the problem of long MRI imaging time. Recently, deep generative models have exhibited great potentials in algorithm robustness and usage flexibility. Nevertheless, none of existing schemes can be learned from or employed to the k-space measurement directly. Furthermore, how do the deep generative models work well in hybrid domain is also worth being investigated. In this work, by taking advantage of the deep energy-based models, we propose a k-space and image domain collaborative generative model to comprehensively estimate the MR data from under-sampled measurement. Equipped with parallel and sequential orders, experimental comparisons with the state-of-the-arts demonstrated that they involve less error in reconstruction accuracy and are more stable under different acceleration factors.**

# 1. Introduction

Magnetic resonance imaging (MRI) is a leading diagnostic tool that can be used for a variety of diseases, including musculoskeletal, nervous system, and tumor diseases. Unfortunately, the physics characteristics of the data acquisition process make it inherently slower than alternative modalities like CT or X-Ray. Lengthy acquisition times makes MR less patient friendly and increases the per patient examination cost. Consequently, increasing the speed of MRI acquisition has been a major ongoing research goal for decades.

Over the years, several methods have been proposed to accelerate MRI. One of the early approaches is the parallel MRI (pMRI) (Griswold *et al.,* 2005), which exploits spatial sensitivity of multiple coils in conjunction with gradient encoding to reduce the data samples that are required for reconstruction. Historically, parallel imaging methods are put in two categories: Approaches that operated in the image domain, inspired by the sensitivity encoding (SENSE) method (Pruessmann *et al.,* 1999), and approaches that operated in k-space domain, inspired by the simultaneous acquisition of spatial harmonics (SMASH) (Sodickson *et al.,* 1997) and generalized auto-calibrating partial parallel acquisition (GRAPPA) (Griswold *et al.,* 2002). Besides the physical properties of multi-channel acquisition, many endeavors have been made to use the signal properties in MR image reconstruction. Among them, compressive sensing (CS) (Lustig *et al.,* 2008) has represented a widely used accelerating approach, which does not require MR hardware modification. Traditional CS reconstruction techniques are iterative algorithms that usually require a sparsifying transform. These methods have explored wavelet (Chaâri *et al.,* 2011), total variation (Ye *et al.,* 2011), dictionary learning (Liu *et al.,* 2013) to trigger either sparsity on the original image, or joint sparsity such as L1-SPiRiT (Lustig *et al.,* 2009) across the multi-channels. When combined with regularization parameters, they are able to find a solution for these ill-posed inverse problems. Later, the concept of low-rank modeling has shown an increased popularity, such as the eigenvector-based iterative self-consistent parallel imaging reconstruction (ESPiRiT) (Uecker *et al.,* 2014), simultaneous autocalibrating and k-space estimation (SAKE) (Shin *et al.,* 2014), low-rank matrix modeling of local k-space neighborhoods (LORAKS) (Haldar *et al.,* 2014) and annihilating filter based low-rank Hankel matrix (ALOHA) (Jin *et al.,* 2016). Nevertheless, many traditional parallel imaging techniques are time-consuming, making them difficult to be incorporated into near real-time MR imaging scenario.

Recent studies have begun to demonstrate that fast MRI algorithms based on deep learning (DL) methodology could provide even greater flexibility without compromised image quality. Generally, DL methods mainly can be categorized as supervised and unsupervised schemes. So far, most physical-based DL-MRI reconstruction methods belong to the category of supervised learning. Specifically, standard supervised learning-based methods utilize data information to train convolutional neural networks (CNNs) for learning the mapping between the under-sampled and fully sampled data pairs in an end-to-end manner. Hence, they are known to be effective when a large training dataset with accurate labels is available. However, these methods are sometimes impractical due to the requirement of a vast amount of training data is not always practical. In view of this defect, some scan-specific learning methods were developed to alleviate this deficiency that with only the measured data itself is used for self-learning. The robust artificial-neural-networks for k-space interpolation (RAKI) method (Akçakaya *et al.*, 2019) offers an improved k-space reconstruction method using scan-specific deep learning that is trained on autocalibration signal (ACS) data. Later, Arefeen *et al.*, (2022) developed a scan-specific model (SPARK) that estimates and corrects k-space errors made when reconstructing accelerated MRI data.

As an alternative to loose the requirement in supervised deep learning scheme that the training and testing processes should be consistent, the unsupervised deep learning methodology, particularly the deep generative models, exhibits more algorithm robustness and task flexibility. For instance, Liu *et al.*, (2020) leveraged a denoising autoencoding (DAE) network as an explicit prior to address the highly under-sampling MR image reconstruction problem. Tezcan *et al.*, (2018) utilized the variational autoencoders (VAE) prior to produce visually high-quality reconstructions. Another promising direction to utilize generative model is taking the linear inverse problems into account, and then a projected gradient descent algorithm (Bora *et al.*, 2017) for effective use of generative adversarial network (GAN) priors has been proposed. Autoregressive generative models like PixelCNN have also been verified in MRI reconstruction (Luo *et al.*, 2020). In addition, a new generative flow (Glow) (Kelkar *et al.*, 2021) which formulated in the latent space of invertible neural network-based generative models was proposed for reconstructing images from undersampled MRI data. It is worth noting that another generative model EBMs has set off an upsurge (Ji *et al.*, 2011). Recently, by using energy function as the generative model, Guan *et al.*, (2021) proposed an EBMRec method for MRI reconstruction. They provided a new insight that EBMs with an implicit iteration generation and no gradient through sampling are important as they control the diversity in likelihood models and the mode collapse in GANs.

Despite all the successes achieved by the aforementioned generative models, it is easy to discover that they principally work in image domain, no existing schemes that can be learned from or employed to the k-space measurement directly. Furthermore, how do the deep generative models work well in hybrid domain is also worth to be investigated. Therefore, by synergistically combining these findings, we propose a novel model termed KI-EBM that consists of a k-space domain generative network and an image domain generative network. The proposed method jointly takes advantage of information absorbed in k-space and image domain, as opposed to other single domain only approaches (LeCun *et al.*, 2015, Jin *et al.*, 2017, Yaman *et al.*, 2020).

The main contributions of the paper can be summarized as follows:

- As far as we know, the generative model in unsupervised learning fashion is employed in k-space interpolation directly to MRI reconstruction for the first time. More precisely, matrix weighting technique is employed in k-space measurement to ensure the effective generative learning.

- A hybrid-domain method that can process the information generated from the k-space and the image domains is presented for high-precision MRI reconstruction. Particularly, carrying out the combination modes of image domain and k-space domain in both parallel and sequential orders is explored.

- In the reconstruction of multi-coil brain MR data, the integrative EBM model is still trained on single coil data, indicating the algorithm robustness and potential task flexibility.

The rest of this paper is presented as follows. Section 2 briefly describes some relevant works on pMRI and particularly the DL-based pMRI. In Section 3, we elaborate the theory of K-EBM and I-EBM, as well as the issue in K-EBM learning. Section 4 presents two collaborative strategies for KI-EBM, i.e., implementing K-EBM and I-EBM in parallel and sequential orders, and forming pKI-EBM and sKI-EBM. Section 5 and Section 6 report the experimental validation and conclusion of the present methods, respectively.

## 2. Preliminaries

In this section, we first formalize the pMRI reconstruction process, and then review some representative deep-learning methods for pMRI reconstruction.

### 2.1. Background on pMRI

Before explaining the method in detail, we formalize the pMRI process at first. According to the types of underlying domains, MR imaging reconstruction methods are divided into the categories that conducted in image domain and k-space domain. Concretely, the data acquisition for multi-coil pMRI can be described as follows:

$$f_c^K = MFI_c + n_c, \ c = 1, 2, \cdots, C \tag{1}$$

where $f_c^K$ denotes the partially observed measurement of the $c$-th coil. $M \in \mathbb{C}^{N \times N}$ is the undersampling mask, $F \in \mathbb{C}^{N \times N}$ represents the normalized full Fourier transform encoding matrix. $I_c$ stands for the $c$-th coil image and $n_c$ is the noise. To estimate the reconstructed image of $C$ coils, the optimization problem can be expressed as:

$$\underset{I}{Min} \sum_{c=1}^{C} \left\| MFI_c - f_c^K \right\|_2^2 + \lambda R_I(I_c) \tag{2}$$

where the first term represents data fidelity. $\lambda$ is a balance parameter that determines the tradeoff between the prior information and the data fidelity term. $R_I(I)$ enforces prior information to improve reconstruction performance. Calibration-free approaches can be developed to tackle Eq. (2). Specially, if sensitivity information $S_c$ is available, only the final solution $I$ needs to be pursued by letting $I_c = S_c I$. Recently representative of reconstruction approach in image domain is the HGGDP proposed by Quan *et al.,* (2021), which leverages the gradient of data density as prior and significantly improves the native NCSN for high diagnostic-quality image reconstruction.

Alternatively, the data acquisition for multi-coil pMRI in k-space can be described directly by the following formulation:

$$f_c^K = MK_c + n_c, \ c = 1, 2, \cdots, C \tag{3}$$

where $K_c$ represents the $c$-th coil k-space data. The pMRI recovery needs to be formulated as the following optimization:

$$\underset{K}{Min} \sum_{c=1}^{C} \| MK_c - f_c^K \|_2^2 + \lambda R_K(\varphi(K_c)) \tag{4}$$

where $\| MK_c - f_c^K \|_2^2$ is the data-fidelity term. The role of the $\varphi$ is to multiply a weight matrix to adjust the value range of the k-space data, and $R_K(\varphi(K_c))$ is the regularization term of $K_c$. SAKE (Shin *et al.,* 2014) is one of the famous free-calibration imaging reconstruction methods in k-space. It formulates the parallel imaging reconstruction as a structured low-rank matrix completion problem and solves it by iteratively enforcing multiple consistencies.

However, no generative model has been explored on k-space directly so far. Here we develop a straight k-space modeling by choosing the energy-based generative model as the backbone.

## 2.2. Review of DL-based PMRI

*Learning in Image Domain:* A large number of DL-based approaches in image domain have been proposed to tackle the MRI reconstruction problem. The seminal work of Wang *et al.,* (2016) proposed a CNN model to perform off-line training and online single-channel reconstruction for undersampled single-channel data. Meanwhile and thereafter, different deep learning approaches have been developed for fast MR imaging. For example, the work of Hammernik *et al.,* (2018) implemented a variational network which embedded a generalized CS concept for accelerated pMRI reconstruction. Then, Schlemper *et al.,* (2017) used a deep cascade of CNNs to accelerate the data acquisition process. Besides, Qin *et al.,* (2018) presented CRNN-MRI for accelerated dynamic MRI reconstruction. Furthermore, Liu *et al.,* (2019) explored CNN-based inversion in the sparsity-promoting denoising module to generalize the sparsity-enforcing operator. Even though some image domain learning models have CNN or a variational term in the network training loss function, the learning portion of these models happened in the image domain. Therefore, we do not classify these models as hybrid. Moreover, these methods have been limited to image domain data, often generated by taking the Fourier transform of magnitude MR images resulting in a k-space with Hermitian symmetry.

*Learning in K-space:* A commonly perceived problem with MRI reconstruction techniques is the loss of high-frequency information, frequency domain models try to mitigate this problem by including a new domain in the k-space based on the capacity of the fast Fourier transform (FFT). For instance, Han *et al.,* (2019) proposed a fully data-driven deep learning algorithm for k-space interpolation. Specifically, the work continued the trend of using low-rank Hankel matrix (Jin *et al.,* 2016) to directly interpolate the missing k-space data. Extensive numerical experiments show that their proposed method consistently outperforms the existing image-domain DL approaches. According to this observation, Akċakaya *et al.,* (2019) proposed a scan-specific model for k-space interpolation that was trained on the autocalibration signal. Meanwhile, Kim *et al.,* (2019) proposed a similar approach, but used a recurrent neural network model named LORAKI. Both of their experiments outperform the model proposed in image domain.

*Learning in Hybrid Strategy:* Hybrid models leverage information presented in k-space and image domains without attempting to learn the domain transform, making the parameter complexity more manageable. A previous study (Souza *et al.*, 2019) proposed a hybrid model, which was trained end-to-end in a supervised manner, and was assessed only on single-coil data. Eo *et al.*, (2018) introduced a hybrid domain model named KIKI-net, which cascaded k-space domain networks with image domain networks interleaved by data consistency layers and the appropriate domain transform. Further research of KIKI-net (Souza *et al.*, 2019) looked at other possible domain configurations for the sub-networks in the cascade and their results showed that starting the cascade with an image domain sub-network may be advantageous. Based on the success of the above methods, in this work we propose a hybrid approach that works with information captured from k-space domain and image domain. The main difference is that we adopt the generative model for prior learning, rather than the traditional supervised learning strategy.

## 3. Theory of K-EBM and I-EBM

In this section, we first introduce the theory of energy-based generative modeling in image domain and k-space domain. Then, the EBM-based regularization in both domains is formulated for MR reconstruction. Furthermore, the weighting strategy for tackling the issue in K-EBM is presented.

### 3.1. Energy-Based Model (EBM)

Suppose that a dataset consisting of independent identically distributed samples $\{x_i\}_{i=1}^{N}$ from an unknown data distribution $p_D(x)$ that defined on the sample space $\mathbb{R}^m$, our goal is to find the parametric approximation $p_\theta(x)$ of the data distribution. The energy function can be defined as a probability distribution via the Boltzmann distribution:

$$p_\theta(x) = \exp(-E_\theta(x))/Z(\theta) \tag{5}$$

where $Z(\theta) = \int \exp(-E_\theta(x))dx$ denotes the partition function. The main goal of EBM is to learn an energy function $E_\theta(x)$ that assigns low energy values to inputs $x$ in the data distribution and high energy values to other inputs. In our work, this energy function is represented by a deep neural network parameterized with weights $\theta$. The nice property of EBM is that we can freely parameterize the energy in any reasonable way, giving it much flexibility and expressive ability.

We want the distribution defined by $E_\theta(x)$ to model the data distribution $p_D(x)$ by minimizing the negative log likelihood of the data:

$$\mathcal{L}_{ML} = \mathbb{E}_{p_D(x)}[-\log p_\theta(x)] \tag{6}$$

where $-\log p_\theta(x) = E_\theta(x) - \log Z(\theta)$. Thus, the derivative of the log-likelihood is:

$$\nabla_\theta \mathcal{L}_{ML} = \mathbb{E}_{x^+ \sim p_D}[\nabla_\theta E_\theta(x^+)] - \mathbb{E}_{x^- \sim p_\theta}[\nabla_\theta E_\theta(x^-)] \tag{7}$$

Evaluating a universal energy function is convenient to the comparison of the relative probability for different inputs. Nevertheless, for most choices of the energy function $E_\theta(x)$, we cannot compute and even rely on the estimation of the partition function $Z(\theta)$. As a result, estimating the normalized densities is intractable and estimating standard maximum likelihood of the parameters $\theta$ is not straightforward. One commonly used strategy is to use Markov Chain Monte Carlo (MCMC) (Christian *et al.*, 2013) sampling to directly estimate the partition function, in which one iteratively updates a candidate configuration, until these configurations converge in distribution to the desired distribution. There are some well-established approximation methods based on MCMC such as random walk, Gibbs sampler (Stuart *et al.*, 1984), which have long mixing times especially if the energy function is complicated.

To improve the mixing time of the sampling procedure, we exploit MCMC with Langevin dynamics so that the distribution $q_\theta(x)$ will approach to the model distribution $p_\theta(x)$, and this procedure generates samples from the distribution defined by the energy function. Additionally, model parameters are updated based on the maximum likelihood estimation, i.e.,

$$\nabla_\theta \mathcal{L}_{ML} = \mathbb{E}_{x^+ \sim p_D}[\nabla_\theta E_\theta(x^+)] - \mathbb{E}_{x^- \sim q_\theta}[\nabla_\theta E_\theta(x^-)] \tag{8}$$

As a gradient-based MCMC method, Langevin dynamics defines an efficient iterative sampling process, which asymptotically produces samples from an energy-based distribution:

$$\tilde{x}^t = \tilde{x}^{t-1} - \frac{\lambda}{2} \nabla_x E_\theta(\tilde{x}^{t-1}) + \omega^t, \omega^t \sim N(0, \lambda) \tag{9}$$

where the iterative procedure Eq. (9) defines a distribution $q_\theta(x)$ such that $\tilde{x}^T \sim q_\theta(x)$. As stated by Welling *et al.*, (2011), the distribution $q_\theta(x)$ of $\tilde{x}^T$ converges to the model distribution $p_\theta(x) \propto \exp(-E_\theta(x))$ when $\lambda \to 0$ and $T \to \infty$. Thus, samples are generated implicitly by the energy function $E_\theta(x)$ as opposed to being explicitly generated by a feedforward network. In summary, the procedure of sample generation can preclude explicit generator models limitations in a certain way.

## 3.2. EBM-Regularized Reconstruction

Generally, the core idea of this study is to use the energy function to capture certain statistical properties of the input data and map it to energy. Since the essence of EBM is to capture dependencies by associating a scalar energy to each configuration of the variables, EBM does not place a restriction on the characteristics and types of the image, which is more flexible to model a more expressive family of probability distributions. Therefore, we attempt to learn the prior information separately for the image domain and k-space domain. Specifically, Eq. (9) could be cast as $\{x = I^t\}_{t=1}^{T}$ and $\{x = K^t\}_{t=1}^{T}$, respectively. Note that $I^t$ involves the prior information in image domain while $K^t$ can be viewed as the k-space prior information. Therefore, $\{I^t\}_{t=1}^{T}$ and $\{K^t\}_{t=1}^{T}$ stand for the prior-enforcing intermediate values generated by I-EBM and K-EBM, respectively.

Subsequently, by assembling the discussions in the above subsections, the final formulation of I-EBM and K-EBM at each iteration of the annealed Langevin dynamics can be separately formulated as follows:

$$\underset{I_c}{Min} \| MFI_c - f_c^I \|_2^2 + \lambda_I \| I_c - I_c^t \|_2^2 \tag{10}$$

$$\underset{K_c}{Min} \| MK_c - f_c^K \|_2^2 + \lambda_K \| K_c - K_c^t \|_2^2 \tag{11}$$

Similar to the conventional iterative methods, we update the solution in a two-step alternative manner, via tackling the data-fidelity term and the regularization term subsequently. Thus, the least-square (LS) minimization of Eq. (10) and Eq. (11) in I-EBM and K-EBM can be separately solved as follows:

$$(F^T M^T MF + \lambda_I) I_c^t = F^T M^T f_c^I + \lambda_I I_c^t \tag{12}$$

$$(M^T M + \lambda_K) K_c^t = M^T f_c^K + \lambda_K K_c^t \tag{13}$$

Let $F \in \mathbb{C}^{N \times N}$ denote the full Fourier encoding matrix which is normalized as $F^T F = 1_N$. Both $FI(v)$ and $K(v)$ stand for the updated value at under-sampled k-space location $v$, and $\Omega$ represents the sampled subset of data, it yields,

$$FI_c(v) = \begin{cases} FI_c^t(v), & v \notin \Omega \\ \dfrac{f_c^K(v) + \lambda_I FI_c^t(v)}{(1 + \lambda_I)}, & v \in \Omega \end{cases}, \quad c = 1, 2, \cdots, C \tag{14}$$

$$K_c(v) = \begin{cases} K_c^t(v), & v \notin \Omega \\ \dfrac{f_c^K(v) + \lambda_K K_c^t(v)}{(1 + \lambda_K)}, & v \in \Omega \end{cases}, \quad c = 1, 2, \cdots, C \tag{15}$$

The final reconstructions in I-EBM and K-EBM are obtained by combining the channels through the square root of the sum of squares (SOS), i.e.,

$$I = \sqrt{\sum_{c=1}^{C}|I_c|^2} \quad \text{or} \quad \sqrt{\sum_{c=1}^{C}|F^T K_c|^2} \tag{16}$$

To summarize, we conduct a comprehensive analysis and description of the algorithm used for MRI reconstruction. A subsequent question is that can we synergistically integrate I-EBM and K-EBM in comprehensive manner, such that different features in image and k-space domains can be captured. In the next section, we provide two collaborative strategies to leverage the reconstruction performance.

### 3.3. Issue in K-EBM Learning

As mentioned above, the original EBM was separately applied to the modeling of prior information in image domain and k-space domain in previous work. We conduct preliminary experiments and find that I-EBM employed in image domain works well, while K-EBM conducted in k-space domain directly fails. An underlying assumption behind the success of I-EBM in image domain is that the amplitude values under different pixel location in image domain are homogeneous and the range of them is very close. Additionally, the pixel feature in neighborhood is very abundant. By contrast, as visualized in K-EBM of Fig. 1(b), the amplitude values under two neighbor pixels are very different and differ each other very significantly in k-space domain. As a result, not only training the objects in the k-space domain is difficult, but also obtaining a huge amount of useful information is problematic.

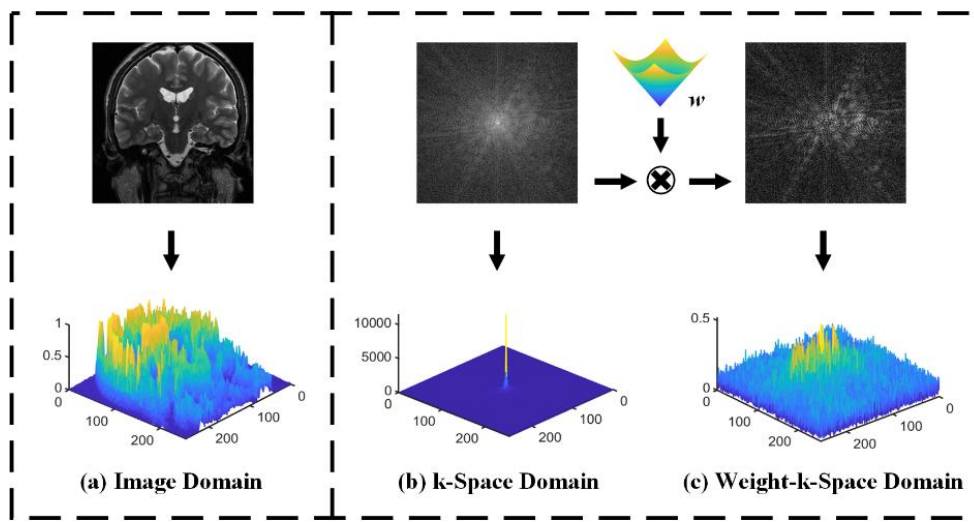

**Fig. 1.** Visual comparison of the amplitude values in different domains. (a) Absolute intensity values of the pixels in mage domain, (b) Amplitude values in the naïve k-space domain, (c) Absolute values in the weight-k-space domain.

Motivated by the works in (Jin *et al.*, 2016, Han *et al.*, 2019, Huang *et al.*, 2008), weighted technique is applied in this study. It can be understood that image support can be efficiently reduced by suppressing low-frequency information with the weighted technique. As visualized in Fig. 1(c), the pixel intensity becomes more uniform, and their range is closer. Subsequently, the introduction of weight technology decreases the amplitude range between pixels even more. We learn EBM model in weighted k-space domain, instead of the naive k-space domain. This manipulation involves two potential benefits: First, the amplitude values of pixel locations in weighted k-space domain are much near and more homogeneous. Second, as well known, the element-wise multiplication in k-space is equal to the convolution operation in image domain. When high-pass filter is employed, the corresponding output is the high-frequency feature which is sparser and consist with the spirit of CS. More rigorous description will be provided in Section 4.

# 4. Collaborative Strategy for KI-EBM

In this section, we mainly introduce two collaborative strategies, coined pKI-EBM and sKI-EBM, to exploit the potentials of K-EBM and I-EBM jointly. They are implemented in parallel and sequential fashions, respectively. Besides, the used network structure in the energy function will be described in detail.

## 4.1. Motivation for Collaborative Learning

**Theorem 1** (Laakom *et al.*, 2021). For the energy function $E(h,x,y) = \frac{1}{2}\|G_W(x) - y\|_2^2$, over the input set $X \subset \mathbb{R}^N$, hypothesis class $H$, and output set $Y \subset \mathbb{R}$, if the feature set $\{\phi_1(\bullet), \cdots, \phi_D(\bullet)\}$ is $\vartheta$-diverse with a probability $\tau$, then with a probability of at least $(1-\delta)\tau$, the following holds for all $h$ in $H$.

$$H = \{G_W(x) = \sum_{i=1}^{D} w_i \phi_i(x) = w^T \Phi(x) \mid \Phi \in F, \forall x \, \|\Phi(x)\|_2 \leq A\} \tag{17}$$

$$\mathbb{E}_{(x,y)\sim D}[E(h,x,y)]\sqrt{DA^2 - \vartheta^2} \leq 8D\|w\|_\infty (\|w\|_\infty \sqrt{DA^2 - \vartheta^2} + B)R_m(F) + \frac{1}{m}\sum_{(x,y)\in S} E(h,x,y) + (\|w\|_\infty \sqrt{DA^2 - \vartheta^2} + B)^2 \sqrt{\frac{\log(2/\delta)}{2m}}$$

$$(18)$$

where $B$ is the upper-bound of $Y$, *i.e.*, $y \leq B, \forall y \in Y$.

As stated in (Laakom *et al.*, 2021), **Theorem 1** provides a generation bound for the EBM learning in regression task. The expected energy is bounded by the sum of three terms: Empirical expectation of energy over the training data, Rademacher complexity of the energy class, and the term involves the number of training data $m$. As can be observed, the bound of the true expectation of the energy decreases with respect to the $\vartheta$-diversity value, i.e., increasing the diversity of the feature set can boost the performance of the EBM model. Besides, the larger the number of training data $m$ is, the lower the bound is. Hence, collecting more training samples also boosts the learning performance.

Based on **Theorem 1**, we can draw the conclusion that the feature diversity and the sample number have significant influence on EBM model. Particularly, increasing the diversity of the feature set can largely boost the generalization performance. In this work, we adopt the diversity strategy of collaborative learning in hybrid domains to leverage the EBM learning. The proposed KI-EBM is formed of two inner-models that learning complementary features from k-space domain and image domain separately to generate energy mapping for each input configuration. These components are described in the following subsections.

## 4.2. Collaborative Learning via Hybrid Schemes

There are two inner-models that play the utmost role in our approach: 1) An EBM model operating in k-space domain (K-EBM); 2) An EBM model operating in image domain (I-EBM). The pipeline of their training process is illustrated in Fig. 2. Each EBM model is trained to minimize the loss between the generative samples and the reference data.

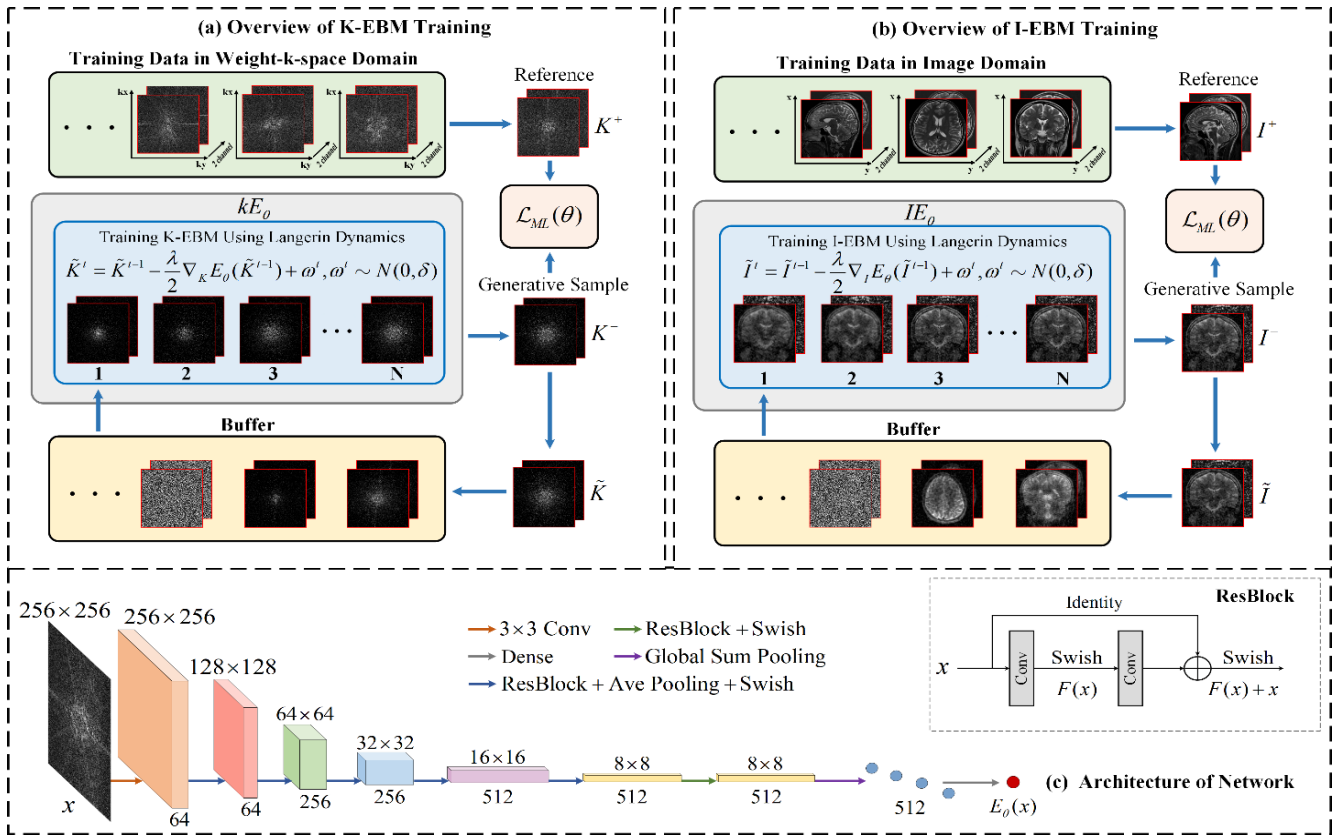

**Fig. 2.** The training flowchart of the proposed KI-EBM. (a) The overview of weight-k-space domain training. (b) The overview of image domain raining. (c) The residual model with multiple residual network architectures and parameters, which is the prior model used in this study. The complex-valued k-space data with size of $256 \times 256$ is split in two-channels: one for the real and other for the imaginary components. Here symbols $K$ and $I$ denote the variables in weight-k-space and image domain, respectively. For more details, please refer to Eq. (6).

On one hand, the training procedure in I-EBM is the same as the work of Guan et al., (2021). On the other hand, as described in Eq. (14) and Section 3.3, training K-EBM directly in the naïve k-space domain is problematic. In fact, central region in the original k-space consists of low frequency information and peripheral region consists of high frequency information. Since the range of low spatial frequency and high spatial frequency varies greatly, training directly on k-space domain is intractable. Thus, we multiply k-space data by weight coefficient as training data. In other words, we multiply or divide the k-space data by a weight coefficient (Jin et al., 2016, Han et al., 2019, Huang et al., 2008) to change the range between high frequency and low frequency. It is further elaborated in mathematical form as:

$$\begin{cases} \varphi_w(K(v)) = w(v) \cdot K(v) \\ K(v) = \varphi_w(K(v))/w(v) \end{cases} \quad (19)$$

$$w(v) = (r \cdot k_x^2 + r \cdot k_y^2))^p \quad (20)$$

where the $w(v)$ can be seen as a high-pass filter in the image domain. $K(v)$ represents the intensity value in the k-space location $v = (k_x, k_y)$ and $\varphi_w(K(v))$ denotes the intensity value with uniform amplitude span. $k_x$ is the count of frequency encoding (FE) lines and $k_y$ is the count of phase encoding (PE) lines. $r$ and $p$ are two parameters to adjust the weight. $r$ sets the cutoff value and $p$ determines the smoothness of the weight boundary. For I-EBM, we use fully sampled MR images as the network input and disturb it simultaneously via random Uniform noise of various amplitudes. During the phase of training, the buffer is used to store past generated samples $\tilde{x}$. Then, we use these samples to initialize Langevin dynamics procedure, which can improve samples generated in the past. In our experiments, we sampled from the buffer 95% of the time, and from uniform noise in the other time. K-EBM and I-EBM capture different features in k-space domain and image domain respectively. Meanwhile, parameter $\theta$ is obtained by maximum likelihood estimation and self-adversarial cogitation, the parameters of the model are optimized to associate the desired configurations with small energy values and the undesired ones with higher energy values.

By selecting the residual network Resnet as an energy function, energy-based distribution representation offers extreme modeling flexibility in the sense that we can use almost model in any domain that performs sample feature learnings. Detailed architecture flowchart and concrete parameter settings in residual model are illustrated in Fig. 2(c). To achieve good non-linearity approximation, Conv layer and ResBlock are adopted in the residual model. In ResBlock, feedforward neural networks with "shortcut connections", which means one or more layers are skipped. More precisely, the shortcut connections simply perform identity mapping, and their outputs are added to the outputs of the stacked layers. Intuitively, one can observe that the filter number in Conv layer of the model is 64 with a kernel size of $3\times3$. Nevertheless, the model under the ResBlock down layer will change the number of filters and increase regularly as the network deepens. Last but not least, the number of filters in the last Global Sum Pooling layer and Dense layer are both 1.

### 4.3. KI-EBM in Parallel and Sequential Orders

The flowcharts of our proposed method for pMRI reconstruction are depicted in Figs. 3-4. Detailed scheme of KI-EBM in parallel order (pKI-EBM) is shown in Fig. 3, and Fig. 4 depicts the whole process of KI-EBM in sequential order (sKI-EBM).

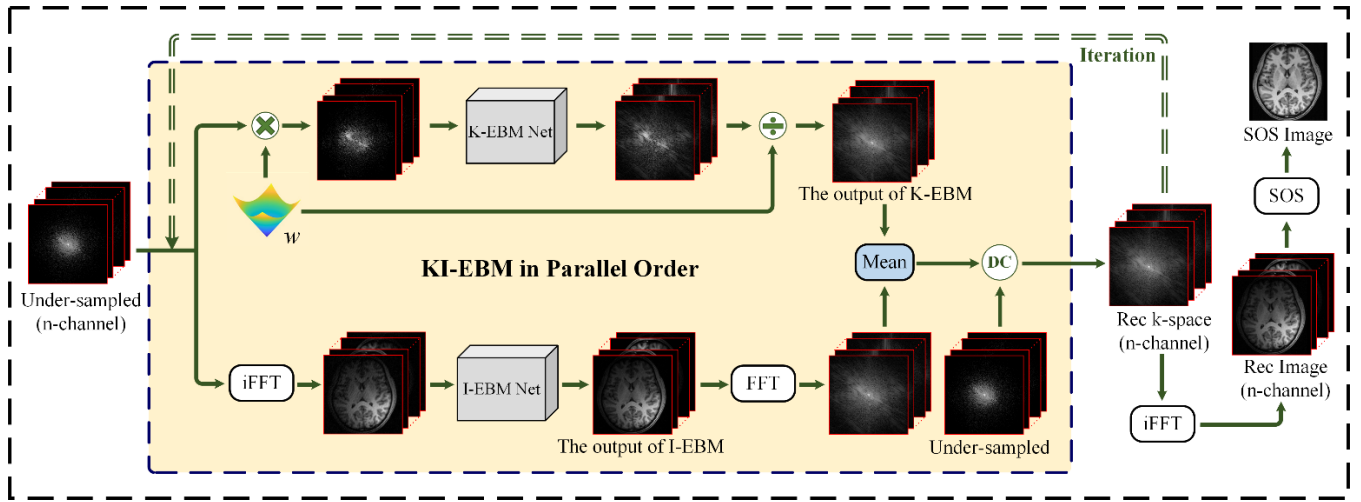

**Fig. 3.** The schema of the proposed pKI-EBM for MRI reconstruction. Note that the images are entering K-EBM and I-EBM at the same time. And we multiply a weight coefficient before the MR image is input to the model, and after K-EBM network output is obtained, it must be divided by the weight coefficient, which is not required during training.

In the case of parallel execution for KI-EBM, we multiply the under-sampled k-space data by the weight coefficient matrix $w$ as the input of the network in K-EBM. At the same time, the intermediate image obtained by applying inverse Fast Fourier transform (iFFT) to the updated k-space data is put into the network in I-EBM. Notice that the network output of I-EBM are images, while the network estimation of K-EBM are k-space data. Therefore, we perform a domain transform on the network output of I-EBM to get k-space data, and average the output of the two networks. Then, we compute the iFFT of k-space data to get multi-channel image. The final reconstruction is obtained by calculating SOS.

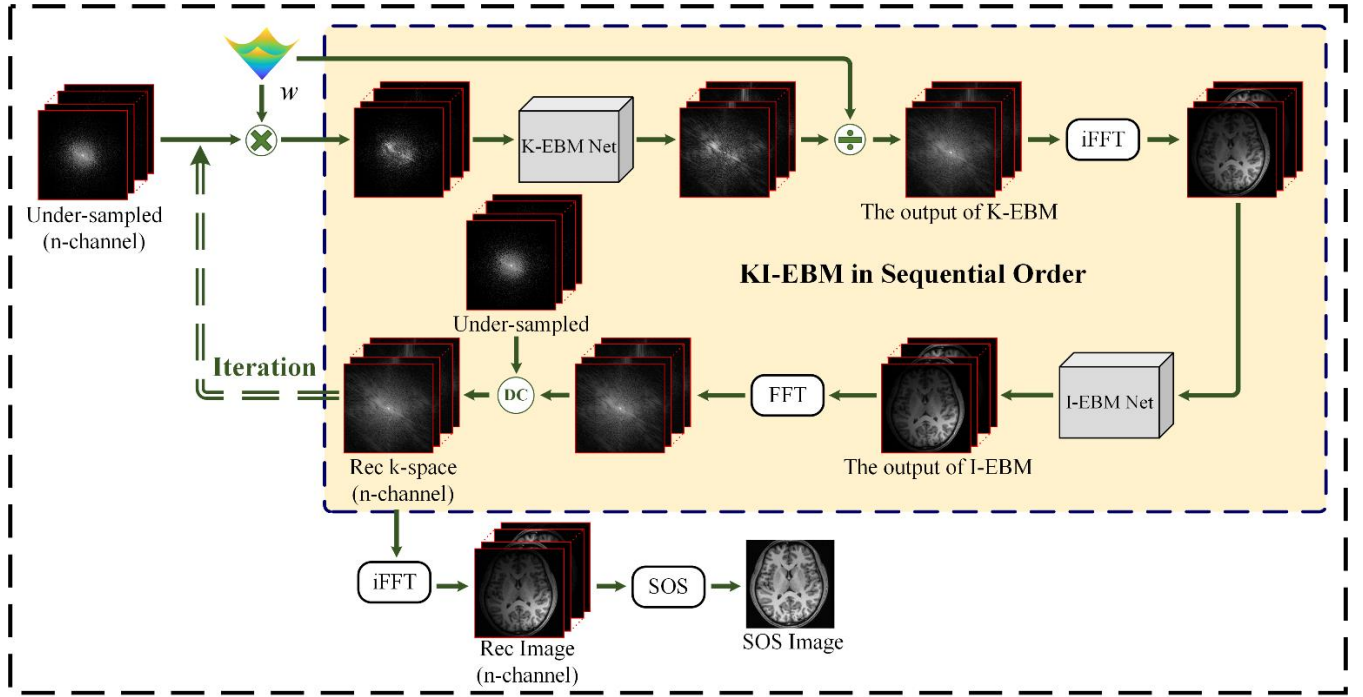

**Fig. 4.** The schema of the proposed sKI-EBM for pMRI reconstruction. Note that the images directly enter I-EBM model after passing through K-EBM model and going through domain transformation.

In the case of implementing KI-EBM in sequential order, we take the under-sampled k-space data multiplied by the weight coefficient $w$ as the input. After passing through K-EBM network, the network output is obtained by dividing the same weight coefficient. Then the output is subjected to iFFT as the network input of I-EBM directly. The frequency domain model can be seen as k-space interpolation to fill the missing values, and the image domain model further improves the intermediate image obtained from the first model.

## 5．Experimental Results

In this section, we introduce the implementation details of the proposed KI-EBM and the used dataset for evaluation. More precisely, the performance of KI-EBM in parallel and sequential orders is validated. Furthermore, we present a set of experiments demonstrating the effectiveness of our method on parallel imaging reconstruction.

### 5.1. Experiment Setup

*Dataset:* The single coil brain images are selected from *SIAT* dataset, which is provided by Shenzhen Institutes of Advanced Technology, the Chinese Academy of Science. Informed consents are obtained from the imaging subject in compliance with the institutional review board policy. The collected dataset includes 500 2D complex-valued MR images with size of $256 \times 256$. In details, MR images are acquired from a healthy volunteer by a 3.0T Siemens Trio Tim MRI scanner using the T2 weighted turbo spin echo sequence. Repetition time (TR)/echo time (TE) is 6100/99 $ms$, and field of view (FOV) is $220 \times 220$ $mm^2$. Besides, the voxel size is $0.9 \times 0.9 \times 0.9$ $mm^3$.

Besides of the single coil dataset, we conduct experiments on multi-coil acquisition data *T1 GE Brain* and *T2 transversal Brain* to verify the performance of KI-EBM for Calibration-free parallel imaging reconstruction. First, the *T1 GE Brain* includes 8-channels complex-valued MR images with size of $220 \times 256$. The MR images are acquired by 3.0T Siemens. The FOV is $220 \times 220$ $mm^2$, and TR/TE is 700/11 $ms$. Second, 12-channels *T2 transversal brain* MR images with size of $256 \times 256$ are acquired with 3.0T Siemens, whose FOV is $220 \times 220$ $mm^2$ and the TR/TE is 5000/91 $ms$.

Finally, experiments are also conducted on multi-coil acquisition data *Test1* of MoDL (Aggarwal *et al.*, 2018) and *Test2* of DDP (Tezcan *et al.*, 2018) to verify the performance of KI-EBM for parallel imaging reconstruction with calibration. Among them, MoDL model is trained on 360 brain MR images (12 coils, complex-valued image). MRI data and Cartesian readouts are acquired using a 3D T2 CUBE sequence and a 12-channel head coil, respectively. The matrix dimensions are $256 \times 232 \times 208$ with 1 $mm$ isotropic resolution. The coil sensitivity maps are estimated from the central k-space regions of each slice using ESPIRiT (Uecker *et al.*, 2014) and are assumed to be known during experiments. Thus, the data has dimensions in rows $\times$ columns $\times$ coils as $256 \times 232 \times 12$. For DDP, it is trained on 790 (single coil, no phase) central T1 weighted slices with 1 $mm$ in-slice resolution from the HCP dataset and tested on an image from a volunteer acquired for this study (15 coils, complex-valued image) along with the corresponding ESPIRiT coil maps.

*Evaluation Metrics:* The commonly used evaluation index Peak Signal to Noise Ratio (PSNR) is included in the reconstruction experiments. In order to avoid the limitation of the PSNR in evaluating image quality, Structural Similarity (SSIM) measurement is also used to evaluate image quality. For the convenience of reproducibility, the source code and some representative results are available at: *https://github.com/yqx7150/KI-EBM*.

*Model Training and Setting:* All networks in the proposed methods are trained using the Adam solver. In the image domain for training I-EBM, the initial learning rate is $3\times10^{-4}$, while the initial learning rate of training K-EBM in k-space domain is $5\times10^{-4}$. The training and respective testing experiments are performed with Tensorflow interface on 2 NVIDIA Tesla V100 GPUs, 16GB.

## 5.2. PMRI Reconstruction with Known Coil Sensitivity

In the circumstance of the coil sensitivity to be known, only the single-coil object needs to be reconstructed. In this subsection, comparison studies are performed with four state-of-the-art parallel imaging reconstruction algorithms, including MoDL (Aggarwal *et al.,* 2018), DDP (Tezcan *et al.,* 2018), HGGDPRec (Quan *et al.,* 2021), and EBMRec (Guan *et al.,* 2021). Quantitative results of different methods are tabulated in Table 1. As well known, flexibility and robustness are the main characteristics of unsupervised learning strategy that differ from the supervised learning counterpart. Notice that the test dataset contains brain MR images, while EBMRec and KI-EBM are trained on the *SIAT* dataset.

**Table 1.** Quantitative PSNR and SSIM measures for different algorithms with varying accelerate factors and sampling masks. (a): results of brain images at various 1D cartesian under-sampling percentages in 15 coils parallel imaging. (b): results of brain images at 2D random sampling in 12 coils parallel imaging.

| (a) | DDP | HGGDPRec | EBMRec | pKI-EBM | sKI-EBM |
|---|---|---|---|---|---|
| R=2 | 37.31 | 39.69 | 39.55 | **40.21** | 39.74 |
| 1D Cartesian | 0.946 | 0.953 | 0.957 | **0.965** | 0.960 |
| R=3 | 33.47 | 36.80 | 37.26 | **38.09** | 37.73 |
| 1D Cartesian | 0.906 | 0.937 | 0.929 | **0.945** | 0.937 |
| (b) | MoDL | HGGDPRec | EBMRec | pKI-EBM | sKI-EBM |
| R=6 | 39.65 | 42.10 | 42.20 | **42.69** | 42.47 |
| 2D Random | 0.936 | 0.972 | 0.989 | **0.990** | 0.989 |

From the metrics in Table 1, it can be observed that the average PSNR values of the reconstructed each image by using KI-EBM are higher than the other models. Particularly, with the increase of acceleration factor, the performance of KI-EBM is much more dominant. This phenomenon further indicates that KI-EBM is more advantageous and effective in severely ill-posed circumstances. To further prove the superiority of KI-EBM, visualization results reconstructed by different methods with different acceleration factors are provided in Figs. 5-6. In general, KI-EBM achieves faithful reconstruction with clear textures and boundaries. For example, compared to DDP and EBMRec reconstruction, the proposed model demonstrates superior reconstructed details that refraining from significant aliasing artifact. Furthermore, KI-EBM is slightly better than HGGDPRec in terms of reconstruction error maps and metrics.

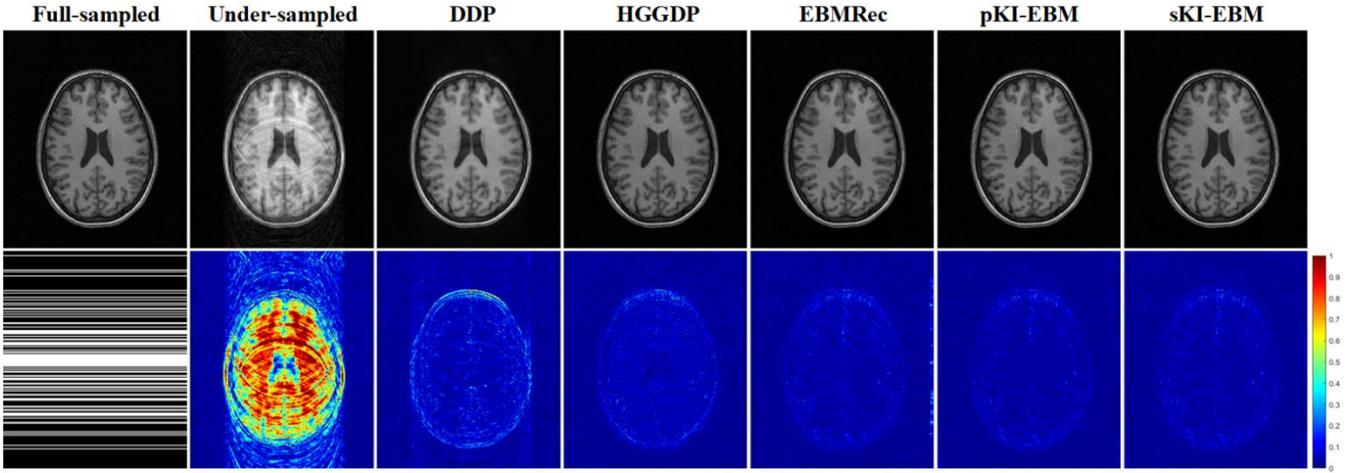

**Fig. 5.** Complex-valued reconstruction results at *R*=3 1D cartesian sampling percentages in 15 coils parallel imaging. From left to right: Full-sampled, Under-sampled, reconstruction by DDP, HGGDP, EBMRec, and KI-EBM.

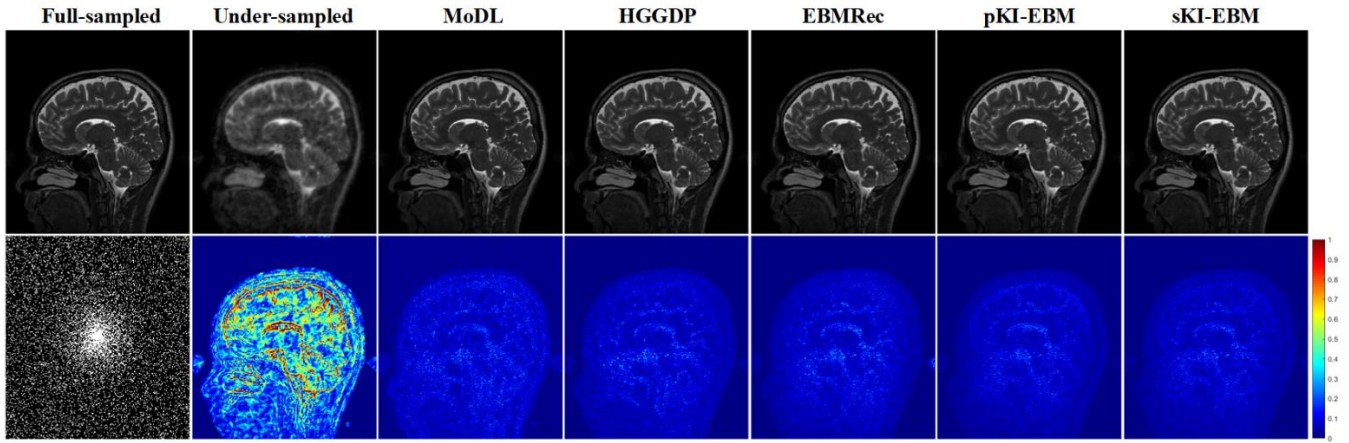

**Fig. 6.** Complex-valued reconstruction results at *R*=6 pseudo radial sampling in 12 coils parallel imaging. From left to right: Full-sampled, Under-sampled, reconstruction by MoDL, HGGDP, EBMRec, and KI-EBM.

## 5.3. Calibration-free PMRI Reconstruction

Calibration-free techniques have presented encouraging performances due to their capability in robustly handling the sensitivity information. To further validate the feasibility of the proposed method, we implement the calibration-free parallel imaging under 2D Poisson and 2D Random disk under-sampling scheme at different acceleration factors. Moreover, several state-of-the-art methods are compared with KI-EBM, including EBMRec (Guan *et al.,* 2021), ESPIRiT (Uecker *et al.,* 2014), SAKE (Shin *et al.,* 2014), and the calibration-free method with joint-sparse codes LINDBERG (Wang *et al.,* 2017). In details, comparison experiments are conducted on different datasets that mentioned in the previous subsection. For fair comparison to LINDBERG, we empirically tune the parameters in the suggested ranges to give their best performances.

Experimental results with varying accelerate factors and different datasets demonstrate the superiority of KI-EBM over state-of-the-arts. Table 2 depicts the quantitative measurement comparison for these methods. For the *T1 GE Brain* dataset, SKI-EBM improves PSNR from 36.55 dB to 38.51 dB under the acceleration factor *R*=6 compared to the image domain method EBMRec. Even in the case of acceleration factor *R*=10 on the *T2 Transversal Brain*, KI-EBM still produces reasonable result. Although KI-EBM is inferior to SAKE when the acceleration factor *R*=4, in the case of *R*=6 KI-EBM outperforms SAKE and produces the highest PSNR and SSIM values, exhibiting a greater advantage at the high under-sampling rate. Additionally, the SSIM values of KI-EBM are closer to 1 when the acceleration factor *R*=4 on the dataset *T1 GE Brain*, which indicates that KI-EBM has good performance in the terms of SSIM metric.

Visual quality of reconstructions for different methods also varies. As can be seen in Figs. 7-8, the reconstructions with LINDBERG at high acceleration rates may create significant aliasing artifacts and lose anatomical details. EBMRec can improve the reconstruction, but it is hard to see a significant improvement when a significant level of aliasing artifact is presented. In comparison, KI-EBM is outstanding in terms of restoring detailed tissue structures that would disappear in other algorithms as well as in simultaneously removing aliasing artifacts. Interestingly, the same conclusion can be reached from different datasets that KI-EBM exhibits the least error.

To show the stability of the proposed method under different domains, the convergence tendency of PSNR curve versus iteration for reconstructing the *T1 GE Brain* is plotted in Fig. 9. Interestingly, it can be seen that the curve is wavy at early iterations and then becomes stable. Although pKI-EBM converges slightly slower than EBMRec (an EBM-based algorithm that operates only on the image domain), the PSNR value of pKI-EBM already exceeds 39 dB at the 25th iteration, while EBMRec reaches only about 36 dB. Therefore, hybrid domain indeed helps the pKI-EBM and sKI-EBM converges more steadily and reaches a higher PSNR.

**Table 2.** PSNR and SSIM comparison with state-of-the-art parallel imaging methods under different under-sampling patterns with varying accelerate factors.

| *T1 GE Brain* (8 coils) | | ESPIRiT | LINDBERG | SAKE | EBMRec | pKI-EBM | sKI-EBM |
|---|---|---|---|---|---|---|---|
| *R*=4 | PSNR | 39.08 | 38.98 | **41.54** | 40.17 | 40.56 | 40.59 |
| 2D Random | SSIM | 0.933 | 0.961 | 0.952 | 0.968 | **0.973** | **0.976** |
| *R*=6 | PSNR | 36.01 | 35.16 | 38.09 | 36.55 | **38.51** | 38.29 |
| 2D Random | SSIM | 0.921 | 0.958 | 0.932 | 0.952 | **0.966** | **0.967** |
| *T2 Transversal Brain* (12 coils) | | ESPIRiT | LINDBERG | SAKE | EBMRec | pKI-EBM | sKI-EBM |
| *R*=4 | PSNR | 31.74 | 32.87 | **33.91** | 33.19 | 33.50 | 33.55 |
| 2D Poisson | SSIM | 0.819 | 0.901 | 0.896 | 0.915 | **0.926** | 0.925 |
| *R*=10 | PSNR | 28.95 | 26.17 | 29.75 | 29.59 | **29.83** | 29.82 |
| 2D Poisson | SSIM | 0.798 | 0.822 | 0.823 | 0.839 | **0.844** | 0.843 |

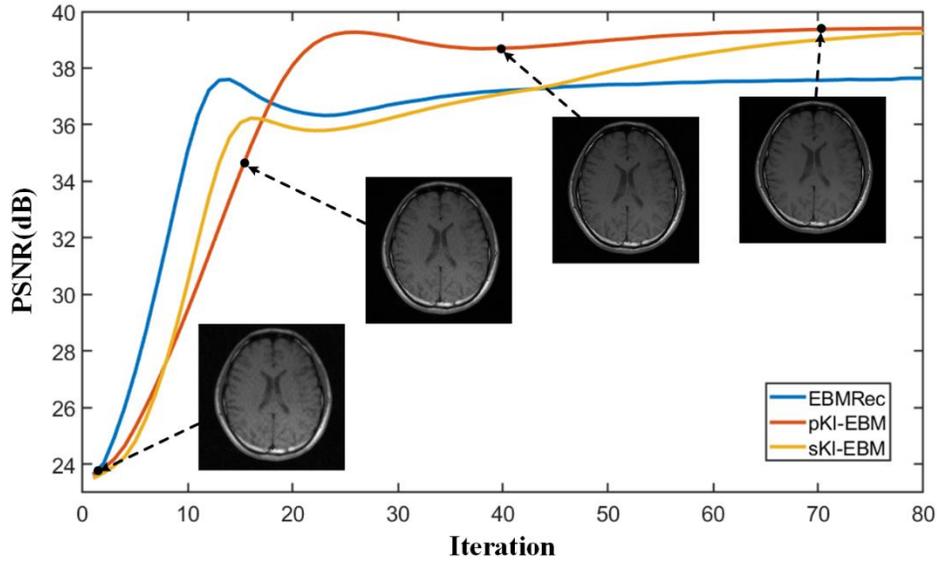

**Fig. 7.** Quantitative PSNR curve versus iteration for reconstructing the *T1 GE Brain* on *R*=5 2D Random sampling pattern. The convergence tendency reflects the stability of the iterative scheme.

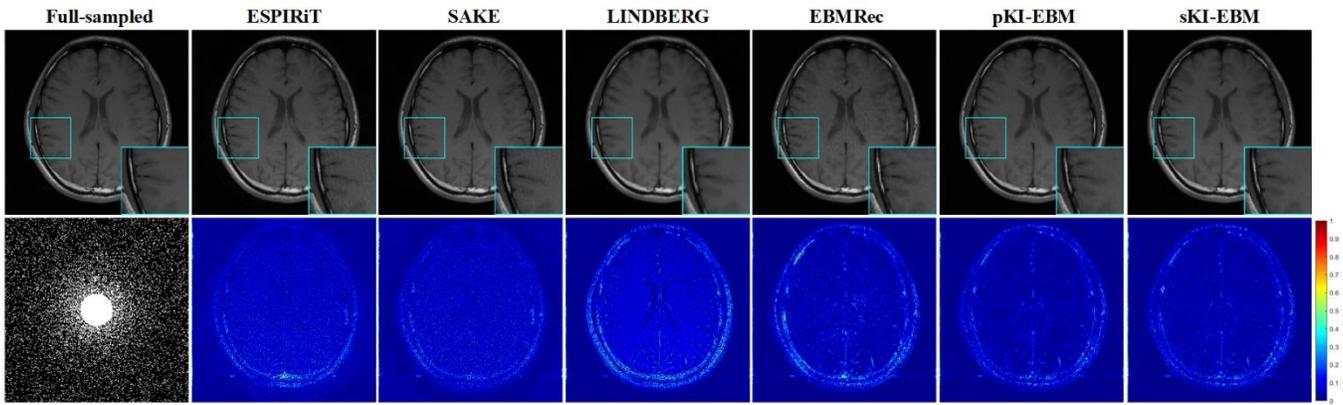

**Fig. 8.** Calibration-free parallel imaging reconstructing results by ESPIRiT, SAKE, LINDBERG, EBMRec, pKI-EBM and sKI-EBM in *T1 GE Brain* at *R*=6 2D Random disk under-sampling mask. The blue box enclosed part has been enlarged to highlight the fine structure information.

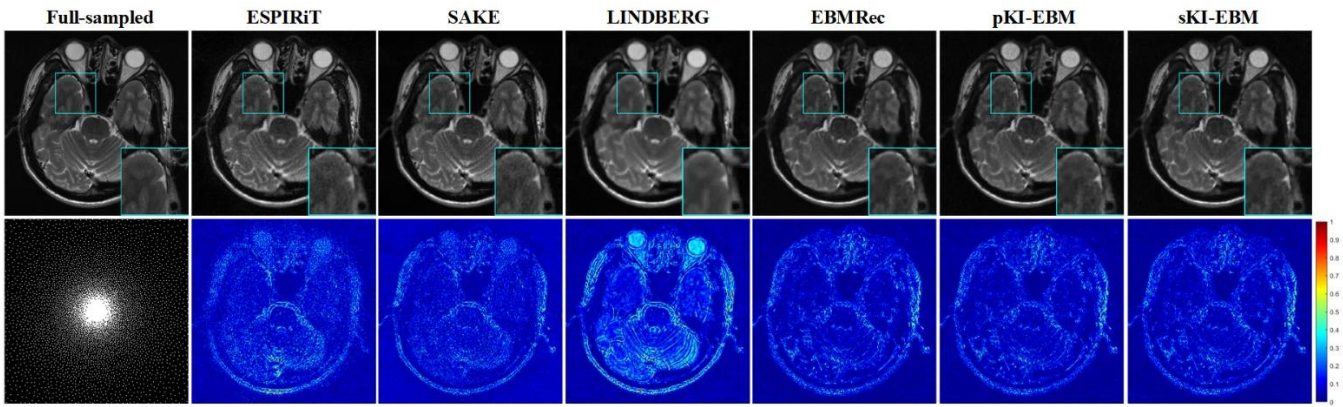

**Fig. 9.** Calibration-free parallel imaging reconstructing results by ESPIRiT, SAKE, LINDBERG, EBMRec, pKI-EBM and sKI-EBM in *T2 Transversal Brain* at *R*=10 2D Poisson disk under-sampling mask. The blue box enclosed part has been enlarged to highlight the fine structure information.

## 5.4. Ablation Study

In this section, we compare the reconstruction results by varying the parameter values of weight. PKI-EBM is trained using five different parameter values of weight as shown in Fig. 10. The two quantitative metrics are listed in Table 3, it can be observed that PSNR and SSIM reach the maximum value when $p = 0.5$ and $r = 1 \times 10^{-1}$. When $p = 0$, no matter what the value of $r$ is, $w$ is a matrix with all elements are 1, which is equivalent to directly train k-space data. When $r = 0$, $w$ is a matrix with all elements are 0. Besides the quantitative comparison, the visual quality is also highlighted in Fig. 11. When $p = 0.5$ and $r = 1 \times 10^{-1}$, the reconstructed image preserves more textural details and has the least noise relative to the reference image. However, the image reconstructed by using the other parameter values of weight does not preserve the detail structures and has the blurring artifacts. To sum up, the parameter values of weight play a significant role in the performance of the proposed KI-EBM, and the reconstruction effect is optimal when $p = 0.5$ and $r = 1 \times 10^{-1}$.

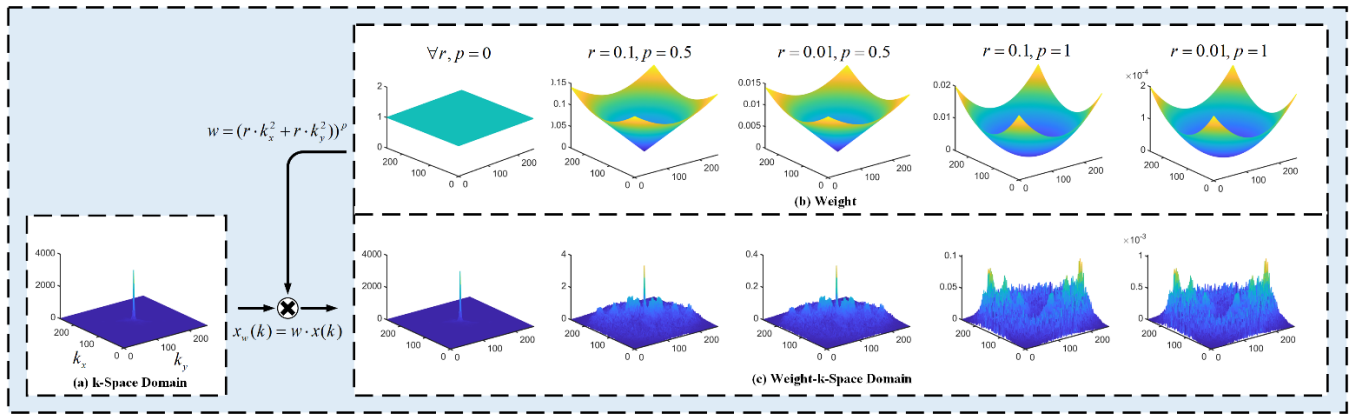

**Fig. 10.** Visual comparison of the amplitude images obtained from the k-space domain and weight-k-space domain under different weight, respectively. (a) The reference k-space data and its amplitude values. (b) The weight with different parameter values. (c) The weight-k-space data and its amplitude values.

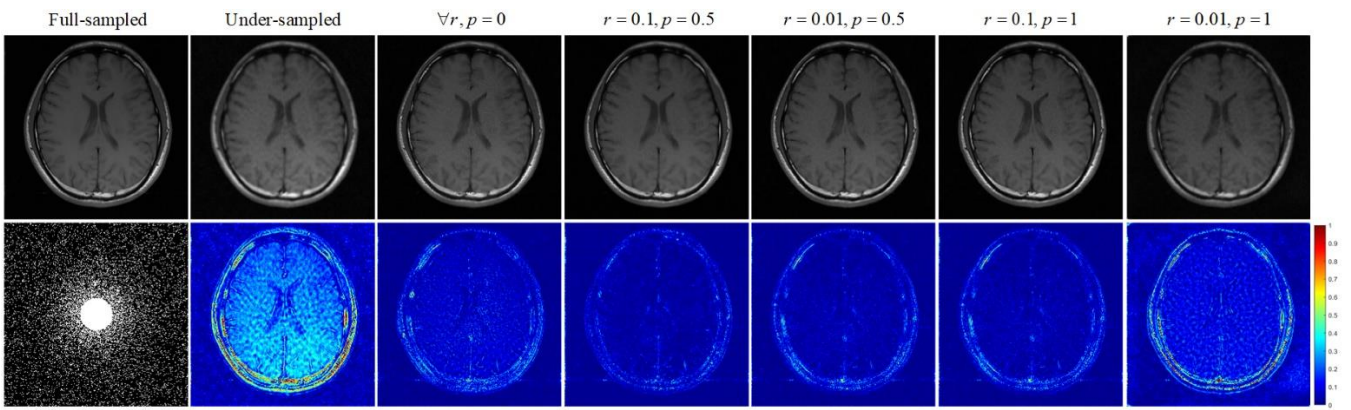

**Fig. 11.** Reconstruction results by pKI-EBM in *T1 GE Brain* at *R*=6 2D Random disk under-sampling mask under different weight. The blue box enclosed part has been enlarged to highlight the fine structure information.

**Table 3.** Quantitative PSNR and SSIM measures for different weight parameter values under $R=6$ 2D Poisson sampling mask for T1 GE Brain.

| pKI-EBM | $p=0$ | $p=0.5$ | $p=1$ |
|---|---|---|---|
| $r=0$ | 34.80/0.950 | -/- | -/- |
| $r=1\times10^{-1}$ | 34.80/0.950 | **38.51/0.966** | 36.88/0.956 |
| $r=1\times10^{-2}$ | 34.80/0.950 | 37.10/0.957 | 30.85/0.831 |

# 6. Conclusion

In this work, a k-space domain and image domain integrative strategy of EBM was proposed for high-quality parallel MRI reconstruction. Particularly, carrying out the combination modes of image domain and k-space domain in parallel and sequential orders was explored. Furthermore, the generative model in unsupervised learning fashion was employed in k-space interpolation directly to MRI reconstruction for the first time by leveraging the matrix weighting technique. With the experimental results discussed above, we have demonstrated that KI-EBM can reliably and consistently recover the nearly aliased-free images with relatively high acceleration factors. In summary, the proposed KI-EBM can obtain a higher PSNR and SSIM, and the details of the reconstructed image are more abundant and more realistic for further clinical scrutinization and diagnostic tasks.

# Acknowledgements

This work was supported in part by National Natural Science Foundation of China under 61871206.